\documentclass[apjl]{emulateapj}
\usepackage{apjfonts} % Times fonts

\usepackage{epsfig}
\usepackage{graphicx}

\newcommand{\ba}{\beta_\textrm{app}}
\newcommand{\n}{\nodata}

\def\bi{\begin{itemize}}
\def\ei{\end{itemize}}
\def\be{\begin{equation}}
\def\ee{\end{equation}}

\def\gtrsim{\mathrel{\hbox{\rlap{\hbox{\lower4pt\hbox{$\sim$}}}\hbox{$>$}}}}
\def\lesssim{\mathrel{\hbox{\rlap{\hbox{\lower4pt\hbox{$\sim$}}}\hbox{$<$}}}}
\def\gtrsim{\mathrel{\hbox{\rlap{\hbox{\lower4pt\hbox{$\sim$}}}\hbox{$>$}}}}
\def\lesssim{\mathrel{\hbox{\rlap{\hbox{\lower4pt\hbox{$\sim$}}}\hbox{$<$}}}}

\shortauthors{LISTER ET AL.}
\shorttitle{VLBA JET SPEEDS OF {\it FERMI} $\gamma$-RAY AGN}
\received{2009 February 12}
\accepted{2009 March 20}
\journalinfo{Astrophysical~journal Letters}
\submitted{}

\begin{document}

\title{A Connection Between Apparent VLBA Jet Speeds and Initial Active Galactic Nucleus 
Detections Made by the {\it Fermi Gamma-ray Observatory}}

\author{
M. L. Lister\altaffilmark{1},
D. C. Homan\altaffilmark{2},
M. Kadler\altaffilmark{3,4,5,6},
K. I. Kellermann\altaffilmark{7},
Y. Y. Kovalev\altaffilmark{8,9},
E. Ros\altaffilmark{8,10},
T. Savolainen\altaffilmark{8},
\\
J. A. Zensus\altaffilmark{8,7}
}

\altaffiltext{1}{
Department of Physics, Purdue University, 525 Northwestern
Avenue, West Lafayette, IN 47907, USA;
\email{mlister@purdue.edu}
}
\altaffiltext{2}{
Department of Physics and Astronomy, Denison University,
Granville, OH 43023, USA;
\email{homand@denison.edu}
}
\altaffiltext{3}{
Dr.\ Remeis-Sternwarte Bamberg, Universit\"at Erlangen-N\"urnberg,
Sternwartstrasse 7, 96049 Bamberg, Germany
}
\altaffiltext{4}{
Erlangen Centre for Astroparticle Physics, Erwin-Rommel Str.~1,
91058 Erlangen, Germany
\email{matthias.kadler@sternwarte.uni-erlangen.de}
}
\altaffiltext{5}{
CRESST/NASA Goddard Space Flight Center, Greenbelt, MD 20771, USA
}
\altaffiltext{6}{
Universities Space Research Association, 10211
Wincopin Circle, Suite 500 Columbia, MD 21044, USA
}
\altaffiltext{7}{
National Radio Astronomy Observatory, 520 Edgemont Road,
Charlottesville, VA 22903-2475, USA;
\email{kkellerm@nrao.edu}
}
\altaffiltext{8}{
Max-Planck-Institut f\"ur Radioastronomie, Auf dem H\"ugel 69, 53121 Bonn, Germany; 
\email{ykovalev,ros,tsavolainen,azensus@mpifr.de}
}
\altaffiltext{9}{
Astro Space Center of Lebedev Physical Institute,
Profsoyuznaya 84/32, 117997 Moscow, Russia
}
\altaffiltext{10}{
Departament d'Astronomia i Astrof\'{\i}sica, Universitat de Val\`encia,
E-46100 Burjassot, Valencia, Spain
}

\begin{abstract}
%250 word limit

In its first three months of operations, the \textit{Fermi Gamma-Ray
Observatory} has detected approximately one quarter of the
radio-flux-limited MOJAVE sample of bright flat-spectrum active galactic nuclei (AGNs) at
energies above $100$ MeV. We have investigated the apparent
parsec-scale jet speeds of 26 MOJAVE AGNs measured by the Very Long
Baseline Array (VLBA) that are
in the LAT bright AGN sample (LBAS).  We find that the $\gamma$-ray
bright quasars have faster jets on average than the non-LBAS quasars,
with a median of 15 $c$, and values ranging up to 34\,$c$. The LBAS
AGNs in which the LAT has detected significant $\gamma$-ray flux
variability generally have faster jets than the nonvariable
ones. These findings are in overall agreement with earlier results
based on nonuniform EGRET data which suggested that $\gamma$-ray
bright AGNs have preferentially higher Doppler boosting factors than
other blazar jets. However, the relatively low LAT detection rates for
the full MOJAVE sample (24\%) and previously known MOJAVE
EGRET-detected blazars (43\%) imply that Doppler boosting is not the
sole factor that determines whether a particular AGN is bright at
$\gamma$-ray energies. The slower apparent jet speeds of LBAS BL Lac
objects and their higher overall LAT detection rate as compared to
quasars suggest that the former are being detected by \textit{Fermi}
because of their higher intrinsic (unbeamed) $\gamma$-ray to radio
luminosity ratios.

\end{abstract}

\keywords{
galaxies: active ---
galaxies: jets ---
radio continuum: galaxies ---
gamma rays: observations ---
quasars: general ---
BL Lacertae objects: general
}
 \ 
%\vfill\eject
%\tableofcontents
 \ 
%\vfill\eject
\section{Introduction}

The $\gamma$-ray emission from highly beamed relativistic jets
associated with active galactic nuclei (AGN) most likely originates
near the base of the jet where the energetics are strongest
\citep{DS94}. High-resolution imaging of this region is therefore critical
for understanding the mechanism by which high energy radiation is
generated. Arguments based on size limits deduced from time
variability and the cross section for pair production suggest that the
$\gamma$-ray emission, like the radio emission, is Doppler boosted
\citep{DS94}. In fact, the $\gamma$-rays may be even more strongly
beamed than the radio emission, since the former generally have a
steeper spectral index $\alpha$ (where $S_\nu \propto \nu^\alpha$),
and the boosting factor is proportional to $\delta^{2-\alpha}$ for
continuous jets. Also, \cite{D95} has shown that if the bulk of the
$\gamma$-rays are produced by external Compton scattering off photons
associated with the accretion disk, then the resulting $\gamma$-ray
emission will be boosted by an additional factor of
$\delta^{1-\alpha}$.

If $\gamma$-ray loud AGNs do indeed have systematically high Doppler
factors, then we might also expect them to have a distinct apparent
speed distribution, since both of these properties depend on the
Lorentz factor and viewing angle of the jet.  Indeed, Monte Carlo
simulations based on a simple linear relationship between radio and
$\gamma$-ray luminosity \citep[e.g.,][]{LM99} confirm that in a
flux-density-limited radio sample, AGN jets with measurable
$\gamma$-ray emission above a fixed sensitivity limit should have
typically higher speeds than those that are not detected in
$\gamma$-rays.

\begin{deluxetable*}{llrl} 
\tablecolumns{4} 
\tabletypesize{\scriptsize} 
\tablewidth{0pt}  
\tablecaption{\label{sampletable}Sample Descriptions}  
\tablehead{\colhead{Sample Name} & \colhead{Description}&
\colhead{Sources} & \colhead{Reference} \\
\colhead{(1)} & \colhead{(2)} &\colhead{(3)}& \colhead{(4)} } 
\startdata 
LBGS & LAT Bright $\gamma$-ray sample & 205 & \cite{Abdo09a} \\
LBAS & LAT Bright AGN List &  106 & \cite{Abdo09b} \\
MOJAVE & MOJAVE radio-selected AGN sample & 135 & \cite{Lister09a} \\
LM   & Intersection of LBAS and MOJAVE lists for $|b|>10^\circ$ & 30 & This Letter
\enddata
\tablecomments{The LBGS covers the entire sky, with higher limiting flux 
in the galactic plane and south celestial pole regions. The LBAS
excludes the sky region $|b|<10^\circ$, while the MOJAVE survey covers
the entire sky north of J2000 declination $-20^\circ$. There are 106 high-confidence AGN associations in the LBAS, and 11 low-confidence assocations listed in \cite{Abdo09b} (one of which also has a high-confidence AGN association with the same $\gamma$-ray source).}
\end{deluxetable*}

Over 65 $\gamma$-ray bright radio sources from the EGRET catalogs
have been identified with flat spectrum radio-loud AGNs (blazars;
\citealt*{MHR01,SRM03,SRM04,CG08}). However, the comparison of radio jet
structure or kinematics with $\gamma$-ray emission has been
inconclusive. \cite{K98} reported no difference in the morphology of
AGN jets with and without observed $\gamma$-ray emission.  However,
using the 2 cm radio data obtained with the NRAO's Very Long Baseline
Array (VLBA)\footnote{The National Radio Astronomy Observatory is a
facility of the National Science Foundation operated under cooperative
agreement by Associated Universities, Inc.}, \cite{KKL05} found that
the radio jets of $\gamma$-ray sources appear to have more compact
structure than non-$\gamma$-ray sources, while \cite{KL04} used
multiepoch 2 cm VLBA data to show that EGRET $\gamma$-ray sources
have marginally higher jet speeds than non-$\gamma$-ray sources.
Also, \cite{J01a} obtained 7 mm and 1.3 cm multiepoch VLBA
observations of 33 $\gamma$-ray blazars between 1993 and 1997 and
reported evidence that AGNs with observed $\gamma$-ray emission have
somewhat higher apparent speeds.
\cite{J01b} claim an association between the time of $\gamma$-ray
flares with the ejection of new superluminal components, and that the
$\gamma$-ray event occurs within the jet features and not at the base
of the jet.  \cite{LV03} found that the $\gamma$-ray flaring preferentially
takes place during the rising or peak period of the high-frequency
radio flares.  These EGRET-based results are consistent with models
where the $\gamma$-ray sources have more highly relativistic jets and
are aligned closer to the line of sight.

However, these earlier findings were based on the somewhat limited and
inhomogeneous $\gamma$-ray data set obtained by the EGRET detector on the
\textit{Compton Gamma-Ray Observatory}, with its small number
statistics, limited sensitivity, and large position uncertainties, as
well as uncertainties introduced by the frequently changing
$\gamma$-ray catalog lists (\citealt*{F94,T95,H99,MHR01,SRM03,SRM04,CG08}).
Thus, it was often ambiguous whether $\gamma$-ray emission was
detected as a result of a flaring event, or was merely the result of a
prescheduled pointed observation, even if the $\gamma$-ray and radio
sources were clearly associated. 

With the beginning of \textit{Fermi} Large Area Telescope (LAT)
operations in 2008 August \citep{Atwood09}, $\gamma$-ray observations
with greatly improved sensitivity and time sampling are now available
for a large sample of AGNs. These data make it possible to recognize
and study $\gamma$-ray flaring activity with a time resolution of
days.  At the same time, the continuation and extension of the 
MOJAVE VLBA program\footnote{http://www.physics.purdue.edu/MOJAVE} is
measuring the velocity and ejection epochs for bright features in over
200 relativistic jets associated with bright radio-loud AGNs
\citep{Lister09a}. 

In this Letter, we discuss the parsec-scale jet kinematic
properties of AGNs in the radio-flux-limited MOJAVE sample that are associated with
bright $\gamma$-ray sources detected during the first three months of LAT
all-sky survey observations. The latter are $>10\sigma$ detections from
the bright $\gamma$-ray source list as reported by
\cite{Abdo09a}. The association of these sources with AGNs (the LAT
bright AGN sample: LBAS) is discussed by \cite{Abdo09b}.   Since the
LBAS is not a uniform, flux-limited list of LAT $\gamma$-ray
detections, the goal of our investigation presented here is mainly to
identify general trends in the preliminary $\gamma$-ray data. A more
detailed and thorough study will be made at the time of the one-year
\textit{Fermi} data release, which will contain a uniform all-sky
catalog and $\gamma$-ray energy spectra of all LAT-detected sources.

Our analysis focuses on the observed MOJAVE jet speeds of the LBAS
sources, their redshift, and optical classifications, and how these
factors influence their observed $\gamma$-ray flux. In an accompanying
{\it Letter} \citep{Kovalev09}, we report on the correlation of
$\gamma$-ray flux with quasi-simultaneous parsec-scale radio flux
density, radio jet compactness, and overall jet activity level during
the three-month initial LAT observation period.  A full report on the
jet kinematics of all 135 sources in the flux-density-limited complete
MOJAVE sample will be presented by \cite{Lister09b}. We use a lambda
cold dtark matter ($\Lambda$CDM) cosmology with
$H_0=71$~km\,s$^{-1}$\,Mpc$^{-1}$, $\Omega_\mathrm{m}=0.27$, and
$\Omega_\Lambda=0.73$.

\section{Comparison of MOJAVE and LAT-detected AGN samples}
\subsection{Sample definitions}

The MOJAVE program \citep{Lister09a} and its predecessor, the VLBA
2\,cm Survey \citep{K98}, have been monitoring the parsec-scale jet
kinematics of the brightest compact extragalactic radio jets in the
northern sky since 1994. The full monitoring sample currently
comprises over 200 AGNs, and includes a statistically complete,
flux-density-limited subsample of all 135 sources above J2000 declination
$-20^\circ$ having VLBA correlated (i.e., compact) flux density above
1.5\,Jy (2\,Jy for sources south of declination $0^\circ$).  To
account for the variable nature of blazars, the flux density limit
applies to a 10-year time period from 1994.0 to 2004.0, i.e., all
AGNs that were known to exceed this limit during this time range are
included. For the purposes of this discussion, we will hereafter refer
to this subsample as the MOJAVE sample.  Details of the extensive
database search that we conducted to ensure the completeness of the
MOJAVE sample are given by \cite{LH05} and \cite{Lister09a}.

The fact that the MOJAVE sample is selected on the basis of
relativistically beamed, compact jet emission at centimeter-wavelengths makes
it less prone to obscuration effects that can bias other
blazar surveys. A detailed study of selection biases by \cite{LM97}
suggests that this type of large, radio-selected sample should contain
many of the fastest jets in the parent population, since the latter
can have very high Doppler boosting factors (up to $\sim2 \Gamma$,
where $\Gamma$ is the bulk flow Lorentz factor) for favorable viewing
angles. This Doppler orientation bias ensures that the MOJAVE sample
is composed primarily of jets having high-$\Gamma$ and low-viewing
angles (i.e., flat-spectrum radio quasars and BL\,Lac objects).  Its
high blazar fraction and large sky area coverage therefore make it a
very useful sample for comparing the radio and $\gamma$-ray properties
of AGN jets.

Although the MOJAVE sample contains 135 AGNs with declinations above
$\delta >\; -20^\circ$, 12 sources lie in the galactic plane region
($|b| < 10^\circ$) that is excluded from the LBAS list. For the
purposes of our analysis, we will omit these AGNs from the MOJAVE
sample. Of the remaining 123 MOJAVE sources, 30 are in the LBAS list
(including one low-confidence source listed by \citealt*{Abdo09b}),
which corresponds to an overall $\gamma$-ray detection rate of
$24\%$. Hereafter, we will refer to these LBAS-MOJAVE sources as the
LM subset (see Table 1 for a summary of sample descriptions). The LM
sources are listed in Table 2, and include 19 out of 94 MOJAVE quasars
(20\,\%), 10 out of 21 BL\,Lac objects (48\,\%), and 1 out of six radio
galaxies (17\,\%). Neither of the two optically unidentified MOJAVE
sources are in the LBAS list. From a $\gamma$-ray sample perspective,
19 out of 41 LBAS quasars (46\,\%) and 10 out of 31 LBAS BL\,Lac
objects (32\,\%) north of declination $-20^\circ$ and with
$|b|>10^\circ$ are members of the MOJAVE sample.  The only LBAS radio
galaxy north of declination $-20^\circ$ is a MOJAVE source
(NGC\,1275).

For 26 of these 30 bright LM sources, we have measured
parsec-scale apparent jet speeds by tracking moving features in VLBA
images over periods of several years. In the case of two sources, no
redshift information is available; another two were either too
compact or show complicated jet kinematics that could not be
adequately measured by MOJAVE.  We have found that in most jets, the
various features typically move with similar but somewhat different
speeds \citep{KL04}. For our analysis, we use the fastest robust speed
measured in a given jet, which we consider to be most representative
of the speed of the underlying jet flow. The speeds in Table~2
represent the fastest feature in each jet whose motion was determined
to be radially outward from the base of the jet and did not show
appreciable acceleration. A few isolated jets had no features that met
these criteria; in these cases a mean speed from a simple accelerating
fit model \citep{HOW02} was used. Full details of the speed
determinations will be described by \cite{Lister09b}.

\subsection{The Quasar and BL\,Lac populations}

\begin{figure}[t]
\centering
\includegraphics[angle=0,width=0.48\textwidth]{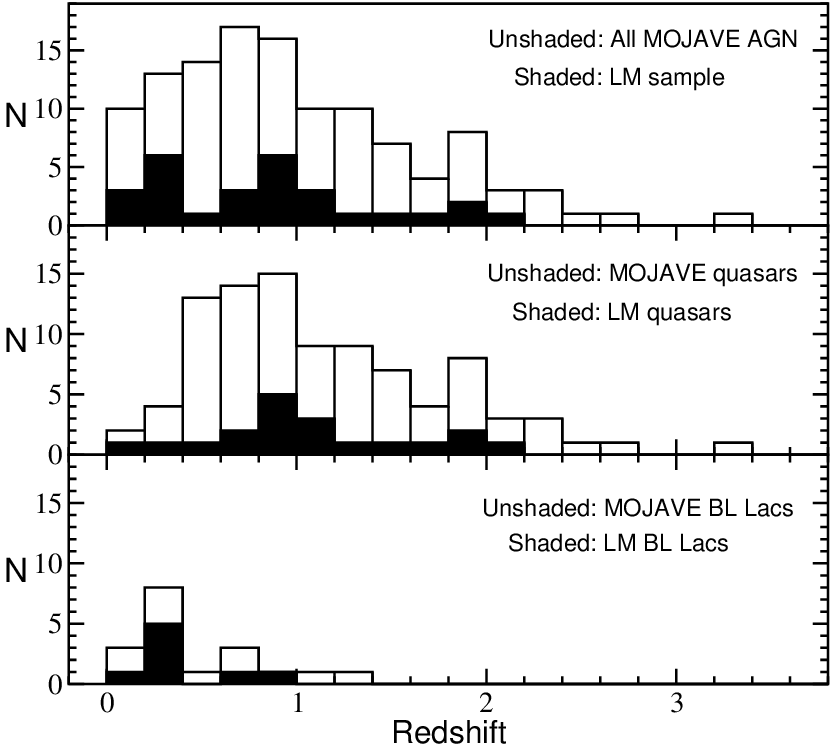}
\caption{\label{fig:z-distributions} Redshift distributions of AGNs in the
MOJAVE sample (unshaded).  The $\gamma$-ray detected (LM) sources in
each bin are shaded. The top, middle, and bottom panels show the
$z$-distributions for the full MOJAVE sample, the quasar subsample
and the BL\,Lac subsample, respectively.}
\end{figure}

Compared to the third EGRET catalog \citep{H99}, the LBAS has a
relatively higher ratio of BL\,Lac objects to quasars. This effect
is understood to be the result of a higher detection efficiency of the
LAT for sources with harder $\gamma$-ray spectra, which is a known
characteristic of BL Lac objects \citep{Abdo09a}.  Moreover, the
redshift distributions of LBAS quasars and BL\,Lac objects show two
well-separated peaks, with the BL\,Lacs peaking at small redshifts
($z<0.5$) and the quasars peaking around $z=1$.  The radio-selected
MOJAVE sample, on the other hand, is dominated by quasars. In light of
possible population differences, we separate the BL\,Lacs and quasars
in our subsequent analysis, noting the better number statistics for
the quasars in our sample.

Figure~\ref{fig:z-distributions} shows the redshift distributions for
the MOJAVE sample separated by optical class, with the LM sources
shaded. In agreement with the overall LBAS results, the LM BL\,Lacs
tend to dominate the low-redshift part of the distributions, while the
distribution of the LM quasars has a maximum close to redshift
$z=1$. A Kolmogorov-Smirnov (K-S) test shows no significant difference
between the redshift distributions of the LM and non-LAT-detected
MOJAVE AGNs. 

\subsection{Apparent parsec-scale jet speeds and $\gamma$-ray variability} 

%\subsection{VLBA Jet Speeds of Bright LAT Blazars}

\begin{figure}[t]
\centering
\includegraphics[angle=0,width=0.48\textwidth,clip]{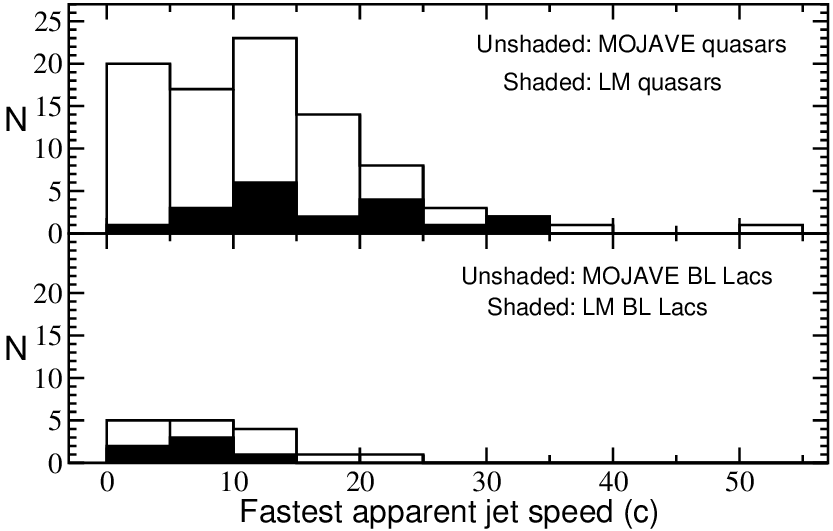}
\caption{\label{fig:latspeeds}
Top panel: maximum jet speed distributions for all MOJAVE quasars. The $\gamma$-ray
detected (LM) quasars in each bin are shaded. Lower panel: same plot for BL Lac
objects. }
\end{figure}

In Figure~\ref{fig:latspeeds} (top panel) we show the distribution of
the fastest jet speeds of the 26 LM sources with MOJAVE-measured
kinematics. There is a significant difference in the speed
distributions of LAT-detected and nondetected MOJAVE quasars. A K-S
test gives a probability of only 2.7\, \% that both distributions are
drawn from the same parent distribution. The LM quasars have higher
VLBA jet speeds, with a peak in the distribution near $\ba \sim
10-15\,c$. Only $5\,\%$ of the slowest MOJAVE quasar jets with $\ba
<5$ are in the LM list. For sources between $\sim 5c$ and $20c$ this
rate increases to $\sim20\,\%$, and above $\ba > 20$ it reaches
50\%. Thus, there is a substantially lower LAT detection rate for the
MOJAVE quasars with the slowest jet speeds.  The statistics for the BL
Lac objects are more sparse, and the K-S test indicates no significant
difference in the LM and non-LM MOJAVE speeds. However, the speed
distributions of the LM BL Lacs and LM quasars are significantly different, 
(0.3\% K-S probability of having the same parent distribution), with
medians of $6c$ and $15c$, respectively. 

\begin{figure}[t]
\includegraphics[angle=0,width=0.48\textwidth,clip]{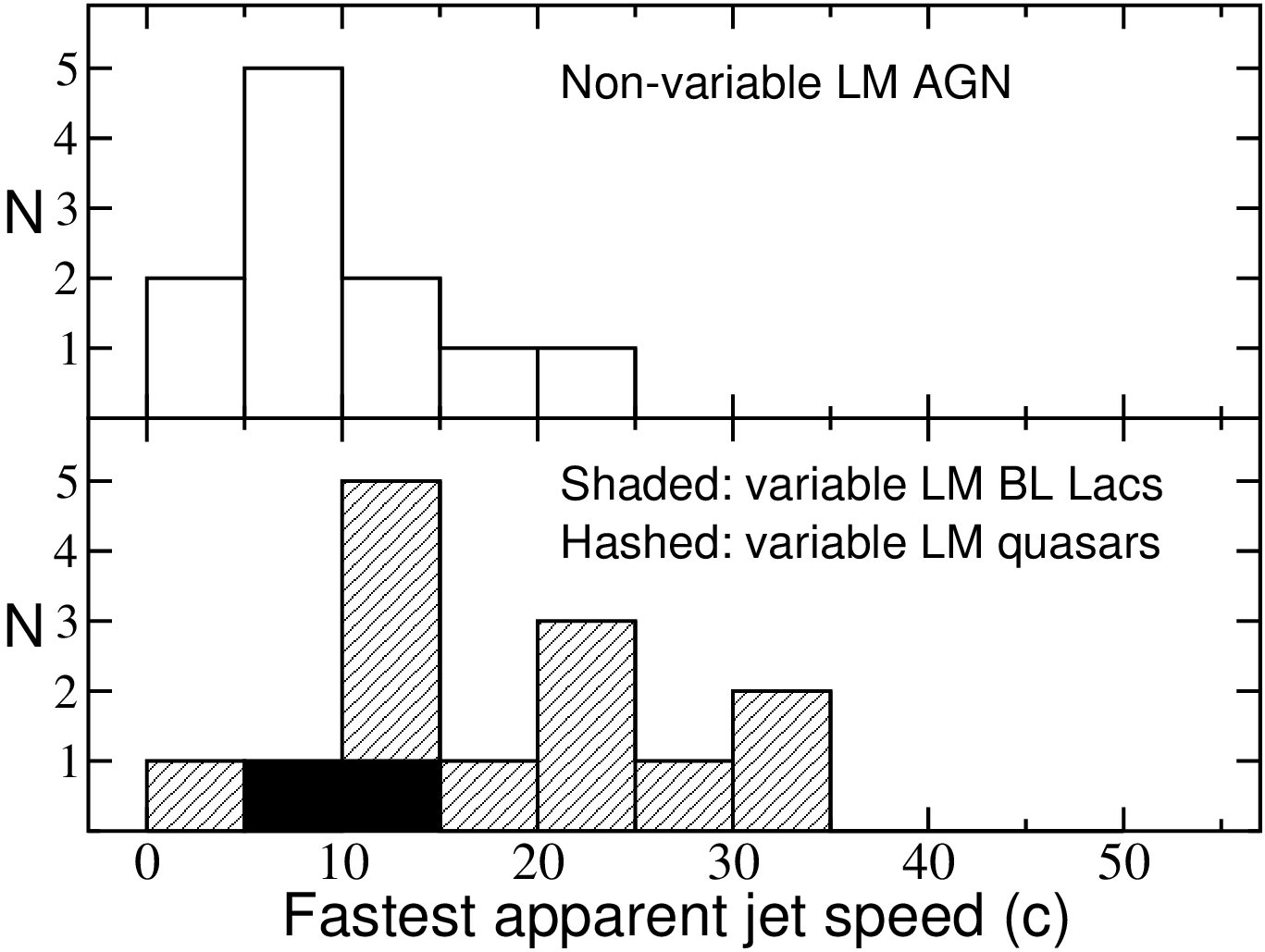}
\caption{\label{fig:latspeedsvariable}
Top panel: apparent jet speed distribution for all LM AGNs
marked as non-$\gamma$-ray variable by \citep{Abdo09b}. 
Bottom panel: apparent jet speed distribution for  $\gamma$-ray variable
BL Lacs (shaded) and quasars (hashed). }
\end{figure}

The LBAS list presented by \cite{Abdo09b} also includes information on
which AGNs have displayed significant $\gamma$-ray variability ($\delta
F/F \gtrsim 0.4$) over the initial three-month LAT observation
period. We find that the $\gamma$-ray variable LM sources exhibit
faster jet speeds (median = $15c$) than the nonvariable ones (median =
$8c$, see Fig.~\ref{fig:latspeedsvariable}). A K-S test gives a
probability of 9.4\% that the speed distributions are similar. If only
quasars and BL Lacs are considered, this value becomes 5\%. The trend
is not present in the quasar-only LM sample.  All but two of the LM
sources with measured VLBA jet speeds of $\beta > 15c$ show evidence
of $\gamma$-ray variability.

\section{Discussion}

A high apparent speed measurement ($\ba$, in units of $c$) in an AGN jet
implies a minimum Lorentz factor of $\Gamma \simeq \ba$ and sets an
upper limit $\theta < 2 \; \textrm{arctan}\; \beta_\textrm{app}^{-1}$
on the viewing angle. The median LBAS quasar speed of $15c$ thus
corresponds to jet viewing angles within $7.6^\circ$ of the line of
sight. Since the observed flaring properties of AGNs are expected to
depend strongly on jet speed and viewing angle \citep{Lister01}, the
trend of $\gamma$-ray variable LBAS sources having even higher speeds
(median = $20c$, $\theta < 5.7^\circ$) is consistent with earlier EGRET
findings (e.g., \citealt*{KL04,KKL05}) that these blazars have
preferentially higher Doppler boosting factors. While large Doppler
boosting factors by themselves may be directly responsible for the
observed correlations, there may be an additional dependence of
intrinsic $\gamma$-ray luminosity on the bulk Lorentz factor of the
jet or rest-frame viewing angle, as both of these quantities are also
closely tied to the apparent superluminal speed.

The relatively low $\gamma$-ray detection rate of the MOJAVE blazar
sample in the first three months of LAT operations (24\%) would indeed
suggest that Doppler boosting is not the sole factor in determining
whether a particular AGN is bright at $\gamma$-ray energies. For
example, only 16 of the 37 MOJAVE AGNs that were high-confidence or
probable EGRET associations according to the references cited in Section~1
are in the LBAS list. Since we would not expect the Doppler factors of
these jets to change significantly in the intervening time period,
this suggests that additional factors are involved. 

In an accompanying Letter, \citep{Kovalev09} we report that the LM
jets tend to be in a relatively active state, as determined by their
jet core compactness and radio luminosity levels. Many AGNs also appear
to have been detected by the LAT by virtue of their proximity (e.g.,
the radio galaxy NGC 1275), or their spectral energy distribution
(i.e., the BL Lac objects).  The slower apparent jet speeds and
fainter VLBA radio luminosities of the BL Lac objects suggest that
they must have higher intrinsic (unbeamed) $\gamma$-ray to radio
luminosity ratios than other blazars, in order to account for their
higher LAT-detection rate. This can be more fully investigated when
the LAT photon energy spectra become available.

\section{\label{summary}Summary}

We have examined the parsec-scale jet kinematics of a subset of 30 AGNs
from the flux-limited MOJAVE VLBA radio sample that were detected
during the first three months of \textit{Fermi} LAT observations at
energies greater than 100 MeV. Although the LAT three-month bright AGN
source list is not a complete flux-limited $\gamma$-ray survey, we
have identified several trends that merit further
investigation with the full one-year \textit{Fermi} data set:

(1) The fraction of MOJAVE AGNs detected by the LAT in its first three
months of operations (30 of 123 = 24\%) is comparable to the fraction
detected by EGRET during its mission lifetime (37 of 123 = 30\%). Only
16 of these AGNs were detected by both $\gamma$-ray telescopes.

(2)  We find no significant difference in the redshift
distributions of the LAT-detected MOJAVE (LM) and non-LAT-detected
MOJAVE AGNs.

(3) The apparent jet speed distribution of the 19 LM quasars with
kinematic information and redshifts is peaked at roughly $10$--$15c$,
while that of the 70 non-LM quasars peaks below $5c$.

(4) Of the 26 LM AGNs with kinematic and redshift information, the 15
that are listed as $\gamma$-ray variable by \cite{Abdo09b} have a
faster median speed ($15c$) than the non-variable ones ($8c$).

(5) The LM BL Lacs have lower redshifts and a slower median jet speed
(6 c) than the LM quasars ($15c$), yet their fractional LAT detection
rate is much higher (48\% vs. 20\%). This is likely because they have
higher intrinsic $\gamma$-ray to radio luminosity ratios than the
MOJAVE quasars.

Although these results are generally consistent with earlier findings
based on nonuniform EGRET data that $\gamma$-ray blazars tend to have
preferentially higher Doppler boosting factors, our results taken
together with those of \cite{Kovalev09} further indicate that the
spectral energy shape and current radio jet activity level are also
important factors in determining whether a particular AGN is visible
at $\gamma$-ray energies.

%% Acknowledgements: 

The authors acknowledge the contributions of the MOJAVE team
as well as students at the Max Planck Institute for Radio Astronomy
and Purdue University. We thank Marshall Cohen for helpful comments on
the manuscript. We also thank David Thompson, Julie McEnergy and the
Fermi LAT team for discussions of their plans for publishing their
bright source list and AGN list, and we look forward to future
cooperation with the LAT team.  M.L.L. is supported under NSF grant
AST-0807860 and NASA-Fermi grant NNX08AV67G. D.C.H. is supported by NSF
grant AST-0707693. TS has been also supported in part by the Academy
of Finland grant 120516.  Part of this work was done by Y.Y.K. and T.S.
during their Alexander von Humboldt fellowships at the MPIfR.
The National Radio Astronomy Observatory is a
facility of the National Science Foundation operated under cooperative
agreement by Associated Universities, Inc.

{\it Facilities:} \facility{VLBA, Fermi}

\begin{deluxetable}{llllllr}
\tablecolumns{7} 
\tabletypesize{\scriptsize} 
\tablewidth{0pt}  
\tablecaption{\label{velocitytable}LM Sample of LAT-Detected MOJAVE AGNs}  
\tablehead{\colhead{Source} & \colhead{LBAS Name}& \colhead{Alias} & \colhead{Class} & \colhead{Var.} & \colhead{z} & \colhead{$\beta_\mathrm{app}$}  \\  
\colhead{(1)} & \colhead{(2)} &\colhead{(3)}& \colhead{(4)} &\colhead{(5)}  &\colhead{(6)} &\colhead{(7)} } 
\startdata 
0048$-$097 & 0FGL J0050.5$-$0928 &  & B & Y  & \n & \n   \\ 
0109+224 & 0FGL J0112.1+2247 &  & B & N  & 0.265 & \n   \\ 
0133+476 & 0FGL J0137.1+4751 & DA 55 & Q & Y  & 0.859 & 13.0 $\pm$ 2.5   \\ 
0215+015 & 0FGL J0217.8+0146 & OD 026 & Q & Y  & 1.715 & 34.2 $\pm$ 2.1   \\ 
0234+285\tablenotemark{a} & 0FGL J0238.4+2855 & CTD 20 & Q & N  & 1.207 & 12.27 $\pm$ 0.84   \\ 
0235+164 & 0FGL J0238.6+1636 &  & B & Y  & 0.94 & \n   \\ 
0316+413 & 0FGL J0320.0+4131 & 3C 84 & G & Y  & 0.0176 & 0.311 $\pm$ 0.059   \\ 
0420$-$014\tablenotemark{a} & 0FGL J0423.1$-$0112 &  & Q & N  & 0.914 & 7.35 $\pm$ 0.98   \\ 
0528+134 & 0FGL J0531.0+1331 &  & Q & Y  & 2.07 & 19.20 $\pm$ 0.42   \\ 
0716+714 & 0FGL J0722.0+7120 &  & B & Y  & 0.31 & 10.07 $\pm$ 0.35   \\ 
0735+178 & 0FGL J0738.2+1738 & OI 158 & B & N  & \n & \n   \\ 
0814+425 & 0FGL J0818.3+4222 & OJ 425 & B & N  & 0.245 & 1.71 $\pm$ 0.29   \\ 
0851+202 & 0FGL J0855.4+2009 & OJ 287 & B & N  & 0.306 & 5.21 $\pm$ 0.40   \\ 
1055+018 & 0FGL J1057.8+0138 & 4C +01.28 & Q & N  & 0.89 & 8.1 $\pm$ 1.4   \\ 
1127$-$145 & 0FGL J1129.8$-$1443 &  & Q & N  & 1.184 & 14.18 $\pm$ 0.59   \\ 
1156+295 & 0FGL J1159.2+2912 & 4C +29.45 & Q & N  & 0.729 & 24.9 $\pm$ 1.8   \\ 
1226+023 & 0FGL J1229.1+0202 & 3C 273 & Q & Y  & 0.158 & 13.44 $\pm$ 0.43   \\ 
1253$-$055 & 0FGL J1256.1$-$0547 & 3C 279 & Q & Y  & 0.536 & 20.58 $\pm$ 0.79   \\ 
1308+326 & 0FGL J1310.6+3220 &  & Q & Y  & 0.997 & 20.88 $\pm$ 0.68   \\ 
1502+106 & 0FGL J1504.4+1030 & 4C +10.39 & Q & Y  & 1.839 & 14.8 $\pm$ 1.2   \\ 
1510$-$089 & 0FGL J1512.7$-$0905 &  & Q & Y  & 0.36 & 20.2 $\pm$ 1.2   \\ 
1633+382 & 0FGL J1635.2+3809 & 4C +38.41 & Q & Y  & 1.814 & 29.5 $\pm$ 1.6   \\ 
1749+096 & 0FGL J1751.5+0935 & OT 081 & B & Y  & 0.322 & 6.84 $\pm$ 0.78   \\ 
1803+784 & 0FGL J1802.2+7827 &  & B & N  & 0.68 & 9.0 $\pm$ 2.5   \\ 
1849+670 & 0FGL J1849.4+6706 &  & Q & Y  & 0.657 & 30.6 $\pm$ 1.5   \\ 
2200+420 & 0FGL J2202.4+4217 & BL Lac & B & N  & 0.0686 & 4.97 $\pm$ 0.30   \\ 
2201+171 & 0FGL J2203.2+1731 &  & Q & Y  & 1.076 & 1.55 $\pm$ 0.33   \\ 
2227$-$088 & 0FGL J2229.8$-$0829 & PHL 5225 & Q & N  & 1.56 & 8.1 $\pm$ 2.1   \\ 
2230+114 & 0FGL J2232.4+1141 & CTA 102 & Q & N  & 1.037 & 15.41 $\pm$ 0.65   \\ 
2251+158 & 0FGL J2254.0+1609 & 3C 454.3 & Q & Y  & 0.859 & 14.19 $\pm$ 0.79
\enddata 
\tablenotetext{a}{$\; $ Low-confidence association in \cite{Abdo09b}.}
\tablecomments{Columns are as follows: (1) IAU name (B1950.0); (2) LBAS catalog name from \cite{Abdo09b}; (3) alternate name; (4) optical class, where Q = quasar, B = BL Lac, G = radio galaxy; (5) $\gamma$-ray variability flag from \cite{Abdo09b};  (6) redshift as tabulated in \cite{Lister09a}; (7) fastest measured radial, non-accelerating jet speed in units of the speed of light.}
\end{deluxetable} 

\end{document}